\documentclass[11pt,a4paper]{article}
\usepackage{hyperref}
\usepackage{bm}
\usepackage{amsmath,amssymb}
\usepackage{cite}

\usepackage{graphicx}

\numberwithin{figure}{section}

\numberwithin{equation}{section}

\newcommand\ie{\textit{i.e.}\ }
\newcommand\eg{\textit{e.g.}\ }
\newcommand\cf{\textit{cf.}\ }

\newcommand{\aka}{{a.k.a.}\ }

\newcommand{\half}{\tfrac{1}{2}}

\newcommand{\be}{\begin{equation}}
\newcommand{\ee}{\end{equation}}
\newcommand{\bea}{\begin{eqnarray}}
\newcommand{\eea}{\end{eqnarray}}

\newcommand{\ph}{\varphi}
\newcommand{\R}{{\cal R}}
\newcommand{\F}{{\cal F}}
\newcommand{\cO}{{\cal O}}
\newcommand{\D}{{\cal D}}
\newcommand{\tg}{{\tilde g}}
\newcommand{\eps}{\varepsilon}

\begin{document}
\begin{titlepage}

\begin{center}
{\huge \bf  Redundant operators in the exact renormalisation group
and in the $f(R)$ approximation to asymptotic safety}
\end{center}
\vskip1cm


\begin{center}
{\bf Juergen A. Dietz and Tim R. Morris}
\end{center}

\begin{center}
{\it School of Physics and Astronomy,  University of Southampton\\
Highfield, Southampton, SO17 1BJ, U.K.}\\
\vspace*{0.3cm}
{\tt  J.A.Dietz@soton.ac.uk, T.R.Morris@soton.ac.uk}
\end{center}


\abstract{
In this paper we review the definition and properties of redundant operators in the exact renormalisation group. We explain why it is important to require them to be eigenoperators and why generically they appear only as a consequence of symmetries of the particular choice of renormalisation group equations. This clarifies when Newton's constant and or the cosmological constant can be considered inessential. We then apply these ideas to the Local Potential Approximation and approximations of a similar spirit such as the $f(R)$ approximation in the asymptotic safety programme in quantum gravity. We show that these approximations can break down if the fixed point does not support a `vacuum' solution in the appropriate domain: all eigenoperators become redundant and the physical space of perturbations collapses to a point. We show that this is the case for the recently discovered lines of fixed points in the $f(R)$ flow equations.
}



\end{titlepage}

\tableofcontents
\vfill
\newpage

\section{Introduction}
\label{introduction}

One attempted route to a quantum theory of gravity is through the asymptotic safety programme \cite{Weinberg:1980gg,Reuter1,Niedermaier:2006wt,Percacci:2007sz,Litim:2008tt,Reuter:2012id}. Although quantum gravity based on the Einstein-Hilbert action is plagued by ultraviolet infinities that are perturbatively non-renormalisable (implying the need for an infinite number of coupling constants), a sensible theory of quantum gravity might be recovered if there exists a suitable ultraviolet fixed point \cite{Weinberg:1980gg}.

Functional renormalisation group (\aka exact renormalisation group \cite{WilsonKogut}) methods provide the ideal framework to investigate this possibility. It is not possible to solve the full renormalisation group equations exactly however. In a situation such as this, where there are no useful small parameters, one can only proceed by considering model approximations which truncate drastically the infinite dimensional theory space. Typically these are polynomial truncations, \ie where everything is discarded except powers of some suitable local operators up to some maximum degree. Nevertheless,  confidence in the asymptotic safety scenario has grown strongly: one finds from these approximate models that qualitative features persist, in particular the existence of a single non-perturbative fixed point supporting three relevant directions, and numerical values for the renormalisation group eigenvalues are reasonably stable, across many different choices of sets of operators, cutoffs and gauge fixings \cite{Reuter1,Niedermaier:2006wt,Percacci:2007sz,Litim:2008tt,Reuter:2012id,Falls:2013bv}.

An important way to go beyond these approximations is to keep an infinite number of operators. In the cases we will be studying, all positive integer powers of the Ricci scalar curvature $R^n$ are kept. Actually, non-perturbative information in $R$ is also incorporated by projecting the Lagrangian onto a general function $f(R)$. In practice this is achieved by working in Euclidean signature and setting the background metric to that of a $d$ dimensional sphere.\footnote{We review this very briefly in sec. \ref{f(R)}.} The renormalisation group flow equations reduce to partial differential equations for this function, and fixed points $f_*(R)$  satisfy ordinary differential equations \cite{Machado, Codello,
Benedetti,Demmel:2012ub,Demmel:2013myx,Benedetti:2013jk}. We will refer to this approximation as the ``$f(R)$ approximation''.

Recently, we solved by a combination of analytic and numerical methods one of the formulations of these equations \cite{Benedetti}, finding fixed point solutions that are globally well defined on four-spheres, \ie in the range $0\le R<\infty$ \cite{Dietz}. However our solutions at first sight seem dramatically at odds with the standard picture: we found continuous lines of fixed points and renormalisation group eigenvalues that are not discrete but continuous.

We showed that sensible fixed point behaviour could in principle be recovered if we extended the domain to the whole real line, thus analytically continuing to negative curvatures. This is consistent with the viewpoint of truncations to polynomials in $R$ since extending the neighbourhood into negative curvatures then happens automatically, but  we are also insisting that $f_*(R)$ is non-singular for all real $R$. However we then found no such fixed point solutions.

In this paper we uncover a much more direct reason for these apparently conflicting results: related to the behaviour over the domain of four-spheres on which the flow equations were defined. We find that for our fixed points there are no `vacuum' solutions\footnote{Strictly this is a misuse of the term, hence the inverted commas. See subsec. \ref{sec-LPA}.} in this domain. By this we mean that there are no $R>0$ solutions to $E_d(R)=0$, the Euclidean equations of motion for constant curvature following from the fixed point action, \cf \eqref{eofm}.
As we show in sec. \ref{f(R)}, there is a dramatic consequence for this type of approximation: the entire space of eigenoperators becomes redundant, meaning that they just generate field reparametrisations of the fixed point action and have no physical consequences. Identifying all the actions that are equivalent under reparametrisations, the entire theory space collapses to a point.

In contrast, we show by asymptotic analysis that had we found a non-singular solution $f_*(R)$ for all real $R$, then this would have supported a `vacuum' solution, implying in this case that none of the operators are redundant.

We have already found that for the flow equations of \cite{Codello}, which are the starting point for very high order truncations \cite{Falls:2013bv}, no global fixed point solution exists, even if we restrict to the space of four-spheres over which these equations were derived \cite{Dietz}. Unfortunately the partial solution picked out by these truncations \cite{Falls:2013bv, Daniel} also does not support `vacuum' solutions within the applicable domain \cite{Kevin}, implying that all its eigenoperators are redundant so that even for this partial solution the theory space collapses to a point.

The key arguments that lead to these conclusions are very straightforward. The impatient reader can find them all in sec. \ref{f(R)}, followed by sec. \ref{conclusions} where we draw further conclusions.

All this relies however on a careful understanding of the properties of redundant operators \cite{WR}, renamed ``inessential" in ref. \cite{Weinberg:1980gg}. The primary purpose of the preceding sections is to build up this understanding by drawing together a number of disparate results in the literature.
Importantly a redundant operator is defined by two properties: that it generates a reparametrisation \emph{and} that it is an eigenoperator. This is covered in sec. \ref{subsec-redundant}. Ignoring the second property is a potential source of confusion, particularly in quantum gravity (namely in considering when Newton's constant and or the cosmological constant can be considered redundant and eliminated from the theory), as we explain in sec. \ref{subsec-why}.

The details in subsections \ref{generalform} and \ref{nonuniversal} are not necessary for the rest of the paper: they are included only because they make the review of redundant operators complete. For completeness also: we show by analytically continuing our asymptotic analysis \cite{Dietz} that had we found a global fixed point solution to the equations of ref. \cite{Benedetti} extended to all real $R$ then this would also have supported a `vacuum' solution for some real $R$, implying that for this situation, none of the eigenoperators are redundant; we include some remarks on the most recent proposal for $f(R)$ flow equations \cite{Benedetti:2013jk}; and we make some remarks on the validity of $f(R)$-type approximations and reparametrisations around the Einstein-Hilbert action and Gaussian fixed point in subsecs. \ref{sec-EHGB} and \ref{EH2}, relating this to investigations of perturbative (non-)renormalisability and the cosmological constant problem.

 In sec. \ref{examples} we cover some known examples of redundant operators. We emphasise that redundant operators appear in the eigenspectrum in general only through some symmetry of the corresponding renormalisation group flow equations. Again this is key to deciding  when a particular coupling  (\eg Newton's constant or the cosmological constant) can be considered inessential. We introduce tests of whether an eigenoperator is redundant, in particular by checking whether, consistently within the approximation in which we are working, it factorizes on the equations of motion.  The intention is to build confidence about the properties of redundant operators through results already established in the literature, before stepping off towards new results about redundancy especially on the already much less certain ground of quantum gravity. Those first steps are taken in sec. \ref{domain} with a careful discussion of the domain over which the equations of motion test should apply.

Then in sec. \ref{LPA-breakdown} we are ready to draw a simple but important general conclusion: the Local Potential Approximation and approximations of a similar spirit such as the $f(R)$ approximation can break down. They do so if there are no solutions to the equations of motion in the domain over which the fixed point solution and its eigenperturbations are defined. In this case all eigenoperators become redundant, so the physical space of perturbations is empty. The results of sec. \ref{f(R)} are now put fully into context.

\section{Review of  properties of redundant operators}

\subsection{General setup}

In this subsection what we will review is very standard but it will help to fix notation and make things concrete. To be precise, consider the exact renormalisation group flow under the cutoff $k$ \cite{WilsonKogut,wegho}, written in terms of  the Legendre effective action \cite{Nicoll,Wetterich,Morris1,Morris:1998da} where it is commonly known as the effective average action \cite{Wetterich}.   For generality, we will write it for a field $\ph^a$, where
the index $a$ stands for Lorentz or internal indices $\alpha$ but also for the position --or on Fourier transform, momentum-- dependence. Contraction of indices then indicates not only the sum over indices but integration over spatial (momentum) dependence (and appropriate signs if the field is fermionic). We work in Euclidean space of dimension $d$. Using the usual flow parameter, namely renormalisation group `time' $t=\ln (k/\mu)$, where $\mu$ is some arbitrary fixed finite renormalisation scale, it takes the form:
\be
\label{erg}
{\partial\over\partial t} \Gamma[\ph] ={1\over2}
\left[\R_{ab} +{\delta^2 \Gamma\over\delta \ph^a\delta \ph^b}\right]^{-1}\!\!\! {\partial\over\partial t} \R_{ba}\,.
\ee
It is the continuum expression of Kadanoff blocking, the first step in constructing a Wilsonian renormalisation group \cite{WilsonKogut}. The computation is performed in Euclidean signature. The inverse is a matrix inverse and the contraction of the indices in particular indicates
a functional trace over the space-time coordinates.
 $\R$ is some infrared cutoff function written as a momentum dependent effective mass term
 \be
 \label{mom-mass}
 \half\,\ph\cdot \R\cdot \ph \equiv \half\,\ph^a\R_{ab} \ph^b = \half\,\int_{p,q} \ph(p)\, \R(p,k)\,\delta(p+q)\, \ph(q)\,,
 \ee
 where in the last equality we recognise that $\R$ is diagonal in momentum space.
Its purpose is to suppress momentum modes $p<k$. Evidently dependence of $\R$ and thus $\Gamma$ on $t$, should be understood, even though we do not indicate it explicitly. For the purposes of this review we can take the traditional form of cutoff where $\R$ does not itself depend on the action $\Gamma$.

This follows straightforwardly by modifying the partition function using \eqref{mom-mass}:\footnote{Note the change of symbol from `classical field' $\ph$ to bare quantum field $\phi_0$.}
\be
\label{partition0}
{\cal Z}_0[J] = \int {\cal D}\phi_0\ \exp\{-S_0 +\half\,\phi_0\cdot
\R\cdot \phi_0+J\cdot\phi_0\}\,,
\ee
differentiating this with respect to $t$ and rewriting this in
terms of $\exp W[J]$, where $W[J]$ is the generator of connected
diagrams, and from there by Legendre transform $\Gamma[\ph]= W+ J\cdot\ph$, to the flow equation
\eqref{erg} \cite{Wetterich, Morris1,Morris:1998da}.

The next step is the rescaling back to the original size \cite{WilsonKogut} which we can conveniently incorporate by writing all quantities in dimensionless terms, taking into account wave-function renormalisation $Z(t)$. In the new variables, with in particular $\R\mapsto Zk^2\R(p^2/k^2)$, \eqref{erg} can be written in compact form as \cite{Wetterich,Morris1}:
\be
\label{ergs}
\left( {\partial\over\partial t} -d_\ph \Delta_\ph - \Delta_\partial +d\right) \Gamma[\ph] =-\, {\rm tr}\ \left[\R +{\delta^2 \Gamma\over\delta \ph\delta \ph}\right]^{-1}\!\!\!\!\left\{(1-\gamma/2)\R+\R'\right\}\,.
\ee
where prime is differentiation with respect to its argument $p^2/k^2$ and the anomalous dimension $\gamma= d\ln Z/dt$. Here we take the liberty of using the same notation for the scaled variables as with the unscaled variables. The running dimension $d_\ph$ for the scaled field $\ph^a$ is for example given by $d_\ph=\tfrac{1}{2}[d-2+\gamma(t)]$ for scalars. Operating on any vertex in the derivative expansion of $\Gamma$,  $\Delta_\ph = \ph\cdot {\delta\over\delta\ph} = \ph^a {\delta\over\delta\ph^a}$ counts the number of occurrences of the field $\ph$, and $\Delta_\partial$ counts the total number of derivatives, being given by \cite{Morris1}:
\be
\label{ddel}
\Delta_\partial = d + \int_p \ph^\alpha(p)\, p^\mu {\partial\over\partial p^\mu} {\delta\over\delta \ph^\alpha(p)}\,.
\ee

For the most part we will not need the detailed form of the flow equation. We take everything from the left hand side in \eqref{ergs} except the first term, put them on the right hand side, and then write the equation more generically as:
\be
\label{generic}
{\partial\over\partial t} \Gamma[\ph] = \F[\Gamma,\ph]\,,
\ee
where $\F$ stands for some functional of its arguments.

A fixed point under the flow is $\Gamma=\Gamma_*$ such that
\be
\label{fp}
{\partial\over\partial t} \Gamma_*[\ph] = 0\,.
\ee
For infinitesimal perturbations around this, \eqref{generic} becomes linear and factorisable. Thus we have $\Gamma=\Gamma_*+\eps\, \cO[\ph] \exp -\lambda t$ where the integrated operator $\cO$ is an eigenoperator solution of
\be
\label{eigen}
-\lambda\, \cO = {\delta\F\over\delta\Gamma}{\Big|}_*\!\!\cdot\cO\,,
\ee
and $\lambda$ is the Renormalisation Group (RG) eigenvalue. The notation on the right hand side here simply means what one gets by forming the linear perturbation:
\be
\label{diff-action}
{\delta\F\over\delta\Gamma}{\Big|}_*\!\!\cdot\cO := \lim_{\eps\to0} {\F[\Gamma_*+\eps\cO]-\F[\Gamma_*]\over\eps}\,.
\ee
The associated infinitesimal dimensionless coupling $g=\eps\,\exp-\lambda t$ is called relevant, marginal or irrelevant according respectively to whether $\lambda$ is positive, zero or negative. Substituting for $t$, we see that it corresponds to a physical coupling $\tg=\eps\, \mu^\lambda$ of definite mass-dimension $\lambda$. Actually, as we will see later, this classification cannot apply to redundant operators; they form a separate class and can be discarded.

Leaving aside also the well-understood exceptions of marginally (ir)relevant couplings (which it will not be necessary to treat here) and exactly marginal couplings, the continuum field theory is completely specified by setting all irrelevant couplings to zero and picking \emph{finite} choices of the ($t$-invariant) values $\eps$ (and thus $\tg$) for all the relevant couplings, recognising that this actually corresponds to an infinitesimal perturbation from the fixed point (\ie infinitesimal scaled couplings) as $t\to\infty$. For these given choices, the whole line parametrised by $t$ is called a renormalised trajectory, and the continuum Legendre effective action is recovered in the $t\to-\infty$ limit \cite{Morris:1998da}. The continuum theory is effectively renormalisable (though not necessarily perturbatively) providing only that the number of relevant couplings is finite.

Later we will deal with flow equations in which $t$-derivatives appear in a more complicated manner than \eqref{generic}, as a result of the cutoff profile itself depending on the effective action. Although the RG eigenvalue equation is then no longer as simple as \eqref{eigen}, none of this affects the properties  of redundant operators that we will need.

\subsection{Definition of a redundant operator}
\label{subsec-redundant}

A redundant operator (renamed inessential in \cite{Weinberg:1980gg}) has a precise meaning \cite{WR}:
\begin{itemize}
\item it is an eigenoperator, and
\item it is equivalent to an infinitesimal change of field variable.
\end{itemize}
The second property means that we can write the operator in the form:
\be
\label{redundant}
\cO = {\delta \Gamma\over\delta\ph}\cdot F\,,
\ee
corresponding to the infinitesimal change of field variable
\be
\label{reparam}
\ph^a\mapsto \ph^a+\eps\, F^a[\ph]\,,
\ee
$F$  being in particular some function of position $x$ and (not
necessarily local) functional of $\ph$, whilst the first property
means in particular that $\Gamma=\Gamma_*$ (but more than this as
discussed in the next subsection).

Actually some comment should be made about this. The natural arena
where changes of field variable should be discussed is in the
partition function \eqref{partition0}. Let $\phi^a_0 \mapsto
\phi^a_0+\eps\theta^a_0[\phi_0]$. After taking into account the
Jacobian, the partition function becomes
\be
\label{nochange}
{\cal Z}_0 = \int {\cal D}\phi_0\ (1+\eps\,\Theta_0) \, \exp\{-S_0
+\half\,\phi_0\cdot \R\cdot \phi_0+J\cdot\phi_0\}\,,
\ee
where
\be
\label{Theta0}
\Theta_0 := \vec{{\delta\over\delta\phi_0^a}}\,\theta_0^a\,,
\ee
the arrow indicating that the derivative acts on everything to its
right. Of course \eqref{nochange} amounts to no change, $\Theta_0$
being a total derivative. If instead we reimpose that the source
couple in a standard way to the (now new) field, then the partition function becomes
\bea
\label{reparam1}
{\cal Z}_0^{(\eps)} &=& \int {\cal D}\ph_0\
\exp\{J\cdot\phi_0\}\,(1+\eps\,\Theta_0) \, \exp\{-S_0
+\half\,\phi_0\cdot \R\cdot \phi_0\}\\
\label{reparam2}
 &=& \int {\cal D}\ph_0\ \exp\left\{-S_0
+\half\,\phi_0\cdot \R\cdot \phi_0 + J\cdot[\phi_0-\eps\,\theta_0] \right\}\\
\label{reparam3}
&=& \left(1-\eps\,J\cdot\theta_0\left[{\delta\over\delta
J}\right]\right) \, {\cal Z}_0\,.
\eea
From \eqref{reparam1} we see that the reparametrisation induces a change in the bare
action and also the form of the cutoff function. The resulting
cutoff function no longer leads to flow of the simple form
\eqref{erg}, but rather one of the generalised types reviewed in
subsec. \ref{generalform}. On the other hand, using integration by parts, we arrive at \eqref{reparam2}, so  it is
equivalent to only coupling the source to the inverse change of
variables. Providing the change is not too non-local this has no
effect on $S$-matrix elements (this being the Equivalence Theorem
\cite{Bergere:1975tr, Itzykson:1980rh}). Using \eqref{reparam3} to rewrite this in terms of
the Legendre effective action (\aka effective average action), by recognising that $J=\delta\Gamma/\delta\ph$, gives
a change of the form \eqref{redundant} with \cite{Morris:1994ie}:
\be
\label{eq-F}
F^a[\ph] =  -\exp(-W[J])\,\theta^a_0\left[{\delta\over\delta
J}\right] \,\exp (W[J])
\ee
(where again the expression on the right is converted to a functional of $\ph$ using $J={\delta\Gamma/\delta\ph}$), thus
justifying the investigation of such changes of variables in the
effective average action in the first place.

The coupling $g$ conjugate to the redundant operator  is called a redundant coupling. Obviously,  to linearised order $\Gamma_*+g\,\cO$ merely reparametrises the action via \eqref{reparam} with $\eps$ replaced by $g$. Intuitively we would expect this also to hold for the full renormalised trajectory generated along this direction. In ref. \cite{WR}, Wegner proves this, order by order in perturbation theory in $g$. The coupling $\tg$ therefore has no consequence on the physics and can be set to zero even if $\lambda>0$. The real set of physical couplings in the continuum theory is therefore the subset of couplings $g$ with $\lambda>0$, having carefully discarded any that are redundant.

\subsection{Why the eigenoperator property is imposed}
\label{subsec-why}

In the literature the first property above is sometimes  ignored or forgotten. Actually, without it one soon runs into difficulties. To see this, consider a fixed point with two eigen-operators $\cO_1$ and $\cO_2$, neither of which is expressible as \eqref{redundant}, both of which are relevant and such that $\lambda_1\ne\lambda_2$. Now suppose that some linear combination is of this form \ie
\be
\label{linear-combo}
\alpha\cO_1+\beta\cO_2 = {\delta \Gamma_*\over\delta\ph}\cdot F_{1,2}
\ee
for some non-zero choice of $\alpha$ and $\beta$ (fixed up to an overall scaling). Any coupling associated to this combination would correspond to an infinitesimal reparametrisation, but the actual couplings $g_1(t)$ and $g_2(t)$ (conjugate to $\cO_1$ and $\cO_2$) grow at different rates $\lambda_i$ under renormalisation group evolution.
Using \eqref{linear-combo} to eliminate a first-order perturbation along $\cO_2$ for example,  would result in an effective conjugate coupling for $\cO_1$ of form $g_1(t) -{\alpha\over\beta}g_2(t)$, but the running of these two bits of the effective conjugate coupling still need to be treated separately. At the non-linear level $g_1$ and $g_2$ generally will mix and evolve entirely differently as the renormalised trajectory develops, further complicating any attempt to force a renormalisation group invariant reduction in the number of relevant couplings.

Precisely this problem arises for the asymptotic safety scenario in the Einstein-Hilbert truncation of quantum gravity
\be
\label{EH}
\Gamma = \int\!\!d^4x \sqrt{g}\,f_{EH}(R)\qquad{\rm where}\qquad f_{EH}(R)={1\over16\pi G}(-R+2\Lambda)\,.
\ee
In this case we have just two dimensions in theory space, given by the Ricci scalar curvature $R$ and cosmological constant $\Lambda$. The combination $R-4\Lambda$ is equivalent to an infinitesimal reparametrisation under a simple rescaling of the metric.\footnote{This led recently to a detailed study \cite{Hamber:2013rb} within some versions of the renormalisation group.} But in practice in the versions the versions of the exact renormalisation group used, neither eigenoperator is this linear combination, so  $R-4\Lambda=\alpha\cO_1+\beta\cO_2$ for some non-vanishing $\alpha$ and $\beta$. There are other subtle features  such as the fact that the rescaling is also equivalent to rescaling the cutoff $k$, and its impact on the resulting strategy for altering the basic flow equation \eqref{erg} to define some appropriate wave function renormalisation. These have been addressed in the literature, see \eg \cite{Percacci:2004sb,Percacci:2004yy,Benedetti:2011ct}. These features are clearly related to the statement we are making and we will have something to say about the latter, but for the moment we want only to
emphasise the reasons why a redundant operator is defined also to be an eigenoperator.

In reality there will always be an infinite number of directions in
theory space. Reparametrisations are then expressible as a sum over
eigenoperators:
\be
\label{general-sum}
{\delta \Gamma\over\delta\ph}\cdot F = \sum_i \alpha_i\, \cO_i\,.
\ee
Generically, infinitely many of the $\alpha_i$ are non-vanishing.
This makes it even less meaningful to attempt a reduction of the
parameter space by `dividing out' by these reparametrisations.

For the Local Potential Approximation and the $f(R)$ approximation in quantum gravity, we will see that
the situation is dramatically different in that it can happen that
all of the eigenoperators are expressible directly as a
reparametrisation, \ie \eqref{general-sum} is satisfied for all eigenoperators with only
one $\alpha_i$ non-vanishing. Nevertheless, this is still
consistent with requiring that to be redundant, an operator must be
both a reparametrisation and an eigen-operator.

In conclusion, we see that  both properties in  subsec. \ref{subsec-redundant} are required in order to allow a renormalisation group invariant elimination of its corresponding coupling.

In subsec. \ref{examples} we will emphasise that as a consequence the very existence of a redundant coupling and its associated operator depends on the type of renormalisation group employed and especially its symmetries.

\subsection{The general form of the Exact Renormalisation Group}
\label{generalform}

Unlike the renormalisation group eigenvalues for all other operators which are universal characteristics of the continuum field theory, the renormalisation group eigenvalues for redundant operators also depend on the choice of the renormalisation group and indeed, by appropriate design of the flow equation, can be chosen at will \cite{WR}. Therefore there is no invariant meaning to the classification in terms of relevant or irrelevant when applied to  redundant operators.

To demonstrate this, we need to take a detour in this and the next subsection. We include these subsections for completeness: we will not need the results of these subsections later. The starting point is to recognise that \eqref{erg} is only one of many forms that the exact renormalisation group can take.\footnote{To keep the discussion general and as simple as possible we return to unscaled variables momentarily. The Legendre transformation in scaled variables have been worked out in ref. \cite{Osborn:2011kw,Rosten:2011mf}.} The other forms do not in general have a simple expression in terms of the flow of the effective average action but are equally valid continuum expressions of the Wilsonian renormalisation group.

The momentum dependent mass term \eqref{mom-mass} means that the massless inverse propagator is infrared regulated and takes the form
\be
\label{Cir}
\Delta^{-1}_{IR}=p^2 + \R(p,k) = p^2/C_{IR}\,.
\ee
Here we are merely  reexpressing the cutoff in multiplicative form as $C_{IR}(p,k)$. It does its job if $C_{IR}\to0$ sufficiently fast as $p/k\to0$, and $C_{IR}\to1$  as $p/k\to\infty$. We will see in a moment why it is useful to introduce a multiplicative ultraviolet cutoff and corresponding propagator $\Delta_{UV}=C_{UV}/p^2$ via:
\be
\label{Cuv}
C_{IR}+C_{UV}=1\,.
\ee
For the moment note that we have (almost) automatically the correct properties: $C_{UV}\to1$ for $p/k\to0$, and $C_{UV}\to0$  sufficiently fast for $p/k\to\infty$.

Defining the interaction part of the effective average action via
\be
\label{Gint}
\Gamma = \Gamma^{int}+ \half \int\!\! d^dx\, (\partial_\mu\ph^\alpha)^2\,,
\ee
it is straightforward to see that \eqref{erg} can be written in the alternative form \cite{Morris1,Morris:1994ie,Morris:1998da}:
\be
\label{leg-flow}
{\partial\over\partial t} \Gamma^{int}[\ph] =-{1\over2}\, {\rm tr}\ \left[1 +\Delta_{IR}\cdot{\delta^2 \Gamma^{int}\over\delta \ph\delta \ph}\right]^{-1}\!\!\! {1\over\Delta_{IR}}{\partial\over\partial t} \Delta_{IR}\,.
\ee
Now define the interaction part of a new action via the Legendre transformation of the effective average action \cite{Morris1}:
\be
\label{leg-trans}
S^{int}[\phi] = \Gamma^{int}[\ph]+\half (\phi-\ph)\cdot\Delta^{-1}_{IR}\cdot(\phi-\ph)\,.
\ee
(Note that the Legendre transform field is written with a different symbol: $\phi$.) The standard Legendre transform identities then become:
\be
\frac{\delta S^{int}}{\delta\phi^{\phantom{int}}} = \Delta^{-1}_{IR}\cdot(\phi-\ph) = \frac{\delta \Gamma^{int}}{\delta\ph^{\phantom{int}}}
\ee
and\footnote{Again, dot indicates contraction of the appropriate free indices, thus
\[
\left(\Delta_{IR}\cdot {\delta\over\delta\phi}\, \ph\cdot\Delta^{-1}_{IR}\right)^a_{\ b} =
{\Delta_{IR}}^{\,ac} {\delta\over\delta\phi^c}\, \ph^d {\Delta^{-1}_{IR}}_{\ db}\,.
\]}
\be
\left[1 +\Delta_{IR}\cdot{\delta^2 \Gamma^{int}\over\delta \ph\delta \ph}\right]^{-1}
= \Delta_{IR}\cdot {\delta\over\delta\phi}\, \ph\cdot\Delta^{-1}_{IR} =1- \Delta_{IR}\cdot {\delta^2 S^{int}\over\delta \phi\delta \phi}\,.
\ee
Differentiating \eqref{leg-trans} with respect to $t$ at constant $\phi$ and substituting \eqref{leg-flow} and the above relations, yields an equation in which $C_{IR}$ appears only in the combination $\partial\Delta_{IR}/\partial t$. Thus,
using \eqref{Cuv}, we arrive at
\be
{\partial\over\partial t} S^{int}[\phi] = {1\over2} \frac{\delta S^{int}}{\delta\phi^{\phantom{int}}} \cdot {\partial\over\partial t} \Delta_{UV} \cdot \frac{\delta S^{int}}{\delta\phi^{\phantom{int}}}
-{1\over2}\, {\rm tr}\ {\partial\over\partial t} \Delta_{UV} \cdot {\delta^2 S^{int}\over\delta \phi\delta \phi}\,.
\ee
This is nothing but Polchinski's version \cite{Polchinski:1983gv} of Wilson's exact renormalisation group \cite{WilsonKogut}. The important point here is to recognise that this is just the functional renormalisation group flow of the effective average action written in a different way. Introducing the total Wilsonian effective action as
\be
S = S^{int} + {1\over2}\phi\cdot\Delta_{UV}^{-1}\cdot\phi\,,
\ee
and scaling to dimensionless variables by writing in particular $C_{UV}\mapsto Z C_{UV}(p^2/k^2)$, the flow equation can be written  similarly to \eqref{erg}:\footnote{Here and in the ensuing, vacuum energy terms are discarded.}
\be
\left( {\partial\over\partial t} -d_\ph \Delta_\ph - \Delta_\partial +d\right) S[\phi] = -\,
\left[{\delta S\over\delta\phi^a}{\delta S\over\delta\phi^b} - {\delta^2 S\over\delta\phi^a\delta\phi^b} -2 (\Delta^{-1}_{UV}\cdot\phi)_a  {\delta S\over\delta\phi^b} \right] (-{\gamma\over2} C_{UV}+C'_{UV})^{ba}\,.
\ee

However, this equation can be written in a more enlightening form as:
\be
\label{flow}
{\partial_t}\, {\rm e}^{-S} = \partial_a \left( \Psi^a {\rm e}^{-S} \right)\,
\ee
where $\partial_t \equiv \partial/\partial t$, $\partial_a \equiv \delta/\delta\phi^a$, and explicitly in momentum space  \cite{Latorre}
\be
\label{Polch-Psi}
\Psi^a[\phi,t] = (d_\phi-d)\phi^\alpha(p) -p^\mu {\partial\over\partial p^\mu}\phi^\alpha(p) -(C'_{UV}-{\gamma\over2} C'_{UV}) \left({\delta S\over\delta \phi^\alpha(-p)}- 2 {p^2\over C_{UV}} \,\phi^\alpha(p)\right)\,.
\ee

We see that actually the functional renormalisation group flow of
the effective average action is, in different variables, nothing
but a particular $t$-dependent change of field variable $\phi^a
\mapsto \phi^a + \Psi^a \delta t$ which thus leaves the effective
partition function\footnote{To keep things simple we are here
ignoring the dependence on the sources. These can be incorporated: see
refs. \cite{Morris1,Morris:1998da}.}
\be
\label{partition}
{\cal Z} = \int {\cal D}\phi\ {\rm e}^{-S}
\ee
invariant, as is immediately clear from \eqref{flow}. The choice \eqref{Polch-Psi} is merely one of an infinite number of choices for $\Psi$ that correspond to different choices of exact renormalisation group \cite{WR,Latorre}.

The transformation properties of the general form of exact
renormalisation group \eqref{flow} under general changes of field
variable were investigated in \cite{WR,Latorre}. They have a very
simple structure if we recognise that \eqref{flow} takes the form
of a one-dimensional gauge theory (with base-space being
renormalisation group time and theory space as fibres)
\cite{Latorre}. Writing $A_t = \vec{\partial_a}\Psi^a$, where as
before the arrow is to emphasise that $\partial_a$ acts on all
terms to its right, \eqref{flow} simply becomes
\be
\label{flow2}
D_t\, {\rm e}^{-S} =0\,,
\ee
where $D_t = \partial_t - A_t$ is a covariant derivative. Consider
a field redefinition $\delta\phi^a=\theta^a[\phi]$ (suppressing
$\theta^a$'s  $t$ dependence) and define similarly to \eqref{Theta0}, $\Theta =
\vec{\partial_a}\, \theta^a$, then the action transforms as in
\eqref{nochange}:
\be
\label{field-redefinition}
\delta\, {\rm e}^{-S} = \Theta\, {\rm e}^{-S}\,,
\ee
and if $\Psi$ transforms via
\be
\delta A_t = [D_t,\Theta]\,,
\ee
then clearly \eqref{flow2} transforms covariantly:
\be
\delta \left( D_t\, {\rm e}^{-S} \right) = \Theta \left( D_t\, {\rm e}^{-S} \right)\,.
\ee

Since $\Psi^a$ and thus $A_t$ are themselves functionals of $S$, see \eg \eqref{Polch-Psi}, fixed points $S^*$ come with a fixed $A^*_t$ which from \eqref{flow2} together satisfy
\be
\label{fixed-point}
D^*_t\, {\rm e}^{-S^*} =-A_t^*\, {\rm e}^{-S^*} = 0\,.
\ee
Perturbing about the fixed point action by some operator with small coupling $g(t)\, \cO$ we have
\bea
\label{eigenoperator}
{\rm e}^{-S} &=& {\rm e}^{-S^*} - g(t)\, {\rm e}^{-S^*} \cO \\
\label{eigenA}
A_t  &=& A^*_t + g(t)\, {\hat \cO}_t\,,
\eea
where ${\hat \cO}_t$ is constructed from ${\hat \Psi^a} = {\delta\Psi^a\over\delta S}{\Big|}_*\!\!\cdot\cO$, and we are using the notation of \eqref{diff-action}. By separation of variables we find $g(t) = \eps \exp -\lambda t$ and arrive at the renormalisation group eigenvalue equation in this framework:
\be
\lambda\, \cO = {\rm e}^{S^*}{\hat \cO}_t \, {\rm e}^{-S^*} - A^*_t \,\cO\,.
\ee

\subsection{The RG eigenvalue of a redundant operator is not universal}
\label{nonuniversal}

Now it is a short step to demonstrate that by appropriate choice of exact renormalisation group, we can choose the renormalisation group eigenvalue of a redundant operator at will \cite{WR}. In fact we will go further and show that  to first order, the entire $t$-dependence of its conjugate coupling $g(t)$ can be chosen at will. Comparing \eqref{eigenoperator} and \eqref{field-redefinition}, we see that a redundant operator can be expressed as
\be
\label{new-redundant}
g(t)\, \cO = -\, {\rm e}^{S^*}  \Theta\, {\rm e}^{-S^*}\,,
\ee
for some generator $\theta$. Of course for a given renormalisation group, such as the one for the effective average action \eqref{Polch-Psi}, $g(t)$ for such a redundant operator will have some fixed scaling dimension \aka RG eigenvalue $\lambda$. On the other hand we have from eqns. \eqref{field-redefinition} -- \eqref{fixed-point}, that if $A_t = A^*_t +[D^*_t,\Theta]$ the equations are satisfied for any choice of $g(t)$. Comparing with \eqref{eigenA}, we see therefore that if we define a new renormalisation group to first order in $g$ via the choice
\be
A^{(g)}_t = A_t - g(t)\, {\hat \cO}_t +[D_t,\Theta]\,,
\ee
this new renormalisation group now has a redundant eigenoperator of the form \eqref{new-redundant} where $g(t)$ has any $t$-dependence we choose.

\subsection{Symmetries and diagnostics}
\label{examples}

Redundant operators are in practice very rare: they appear only in the eigenoperator spectrum if the \emph{flow equation} has an associated symmetry, which importantly means furthermore that their very existence is tied to the choice of flow equation.
The reason is that the RG eigenvalue equation \eqref{eigen} is already constrained to be valid only for discrete values of $\lambda$. Imposing in addition that the operator be expressible in the form \eqref{redundant} then generically over-constrains the equations leading to no solution. In
this subsection we will first explain this in more detail, lay out a detailed example, and then
cover diagnostic tests for recognising when an operator is
redundant. We finish with some other examples in the literature.

In sec. \ref{domain} we address with more care the question of the domain over which the redundancy tests should be applied, and then in sec. \ref{LPA-breakdown} we show that there is one dramatic exception to the findings of this subsection, resulting in all eigenoperators becoming redundant. Finally sec. \ref{f(R)} treats the $f(R)$ approximation to quantum gravity where we will see this break-down in an example.

\subsubsection{The Local Potential Approximation}
\label{sec-LPA}

In the derivative expansion
\cite{Morris:1994ie,Morris:1998da}, the fact that generically there are no eigenoperator solutions to \eqref{redundant} follows from the
parameter counting arguments \cite{Morris:1994jc} recently
re-championed in ref. \cite{Dietz}. Consider for example one-component
 $Z_2$ ($\ph\leftrightarrow-\ph$) symmetric scalar field theory in the LPA (Local
Potential Approximation) in $d>2$ dimensions:\footnote{The special case of $d=2$ dimensions was treated at LPA level in ref. \cite{Morris:1994jc}.}
\be
\label{LPA}
\Gamma[\ph] = \int\!\! d^dx \left\{\half (\partial_\mu\ph)^2 +
V(\ph,t)\right\}\,.
\ee
Here $V$ is the $O(\partial^0)$ part of the effective interactions, viewed as the first approximation in a derivative expansion \cite{Morris:1994ie}. The flow equation with a simple sharp cutoff\,\footnote{Strictly speaking a derivative expansion does not exist in this case \cite{Morris1} but we persist with this choice since it is one of the most studied cases. Different cutoffs result in different right hand sides of the form ${\cal F}(V'')$. Our arguments here hold whatever the exact form.} takes the form
\cite{NCS,Hasenfratz:1985dm,Morris:1994ki,Morris:1998da}:
\be
\label{LPAflow}
{\partial\over\partial t} V(\ph,t) +(1-d/2)\ph V'+d V =
-\ln(1+V'')\,.
\ee
The fixed point potential thus of course satisfies:
\be
\label{LPAFP}
(1-d/2)\ph V'_*+d V_* =
-\ln(1+V''_*)\,.
\ee
Clearly eigen-perturbations around the fixed point potential
$V(\ph,t)=V_*(\ph)+\eps {\rm e}^{-\lambda t} v(\ph)$ therefore
satisfy \cite{Morris:1998da}:
\be
\label{LPAeigen}
\lambda v + (d/2-1)\ph v' -dv = {v''\over1+V''_*}\,.
\ee
For given $\lambda$, this is a linear second-order differential equation and thus requires two boundary conditions to determine a unique solution for $v$. However, we know that $v(\ph)\sim \ph^{(d-\lambda)/(d/2-1)}$ (for $d>2$ dimensions) as $\ph\to\infty$ \cite{Morris:1994jc,Morris:1998da}  and $Z_2$ symmetry imposes $v'(0)=0$ or $v(0)=0$ for even or odd perturbations respectively. Furthermore, linearity allows us to impose a normalisation condition \eg $v(0)=1$ or $v'(0)=1$ respectively. Thus we have three boundary conditions, over-determining the solution space and resulting in quantisation of $\lambda$ \cite{Morris:1994ki,Morris:1994ie,Morris:1998da}.

If we were to further request that the perturbation take the form
\eqref{redundant}, then at the level of the LPA this means that we
want
\be
\label{LPAredundant}
v(\ph) = \zeta(\ph) V'_*(\ph)\,,
\ee
for some non-singular function $F=\zeta(\ph)$ (where $\zeta$ is the
$O(\partial^0)$ part of $F$). This amounts to a diagnostic test for whether the eigenoperator $v$ is redundant. For the Gaussian fixed point solution
to \eqref{LPAFP}, $V_*=0$, there is clearly no solution. As $d$ is
lowered below four, there are non-trivial solutions, characterised
by increasing numbers of turning points and corresponding to higher
order critical points \cite{Codello:2012sc}. We consider the
simplest non-trivial solution (corresponding to the Ising model
universality class). A solution to \eqref{LPAredundant} is possible
if and only if $v(\ph)$ and $V'_*(\ph)$ have the same zeroes. The
explicit solution for $V_*(\ph)$ has two turning points, one at the
origin which is required by symmetry, and one corresponding to a
minimum at some non-zero value $\ph=\pm\ph_*$ (in fact $V_*$ takes
this qualitative form for any cutoff and exact renormalisation
group, see \eg \cite{Morris:1994ie,Morris:1998da,Codello:2012sc}).
For both odd and even operators the condition $v(\ph_*)=0$ is a
further boundary condition on an equation \eqref{LPAeigen} which is
already over-constrained to give quantised eigenvalues. Therefore
generically there are no solutions. (In addition for even operators
there is the further non-trivial condition $v(0)=0$ which
generically cannot be satisfied.) For the multi-critical fixed
points $V_*(\ph)$, \eqref{LPAredundant} is of course even more over-constrained.

The only way out is if there is an exact solution hiding in \eqref{LPAeigen} as a consequence of some underlying symmetry. In fact there is \cite{Morris:1994jc}. Differentiating \eqref{LPAflow} with respect to $\ph$ and setting  $V=V_*$ gives
\[
(d/2-1)\,\ph V''_*-(1+d/2)\, V'_* = {V'''_*\over1+V''_*}\,.
\]
Comparing with \eqref{LPAeigen}, we see that $v=V'_*(\ph)$ is an odd eigenoperator with $\lambda=d/2-1$. It satisfies \eqref{LPAredundant} trivially with $F=1$, and is therefore redundant.

Actually, $F=1$ is an exact solution for the exact renormalisation group \cite{Morris:1994jc}. Substituting this into \eqref{redundant} means that the redundant operator takes the form
\be
\label{exact-redundant}
\cO_r = \int\!\! d^dx {\delta \Gamma_*\over \delta \ph(x)} = {\delta \Gamma_*\over \delta \ph(p)}\Big|_{p=0}\,,
\ee
where the functional derivative is evaluated at $p=0$. Setting $\Gamma=\Gamma_*$ and differentiating \eqref{ergs} with respect to $\ph(0)$, we readily see that \eqref{eigen} is satisfied for $\cO=\cO_r$, providing $\lambda=d/2-1$. The reason we have found this exact solution is an underlying symmetry of the unscaled equation: the fact that $\R$ was chosen to be independent of $\ph$, means that the unscaled flow equation \eqref{erg} is invariant under a constant shift of the field $\ph(x)\mapsto \ph(x)+\delta$.

The symmetry becomes $t$-dependent on scaling to dimensionless variables, which is why the $t$-independent shift $\ph(x)\mapsto \ph(x)+\delta$ is no longer a symmetry of the scaled equations \eqref{ergs} or \eqref{LPAflow}. This is also the reason why the redundant operator has a non-zero RG eigenvalue: a $t$-independent symmetry of the scaled equations would be associated to an exactly marginal ($\lambda=0$) redundant operator, as we will review shortly.

For general approximations or for the exact flow equations, we see from \eqref{redundant} that a simple diagnostic test for whether an operator is redundant is to check  if it vanishes on solutions of the equations of motion
$\delta \Gamma_*/\delta\ph = 0$. We see that this is obviously true of the exact expression for the example redundant operator \eqref{exact-redundant}. From \eqref{LPA}, in the LPA the equation of motion is $V'_*(\ph)=\Box\ph$ which results in \eqref{LPAredundant} vanishing at the LPA, \aka $O(\partial^0)$, level. As we have seen already, in this approximation this means that the only constraints on $v$ being redundant, arise from the points $\ph$ satisfying $V'_*(\ph)=0$. This corresponds to the equations of motion for a constant effective field $\ph$, \ie roughly speaking the vacuum solutions.

{\it N.B.} to find the true vacuum solutions we would first have to complete the construction of the theory by letting $k\to0$. We emphasise that the equations of motion, and the above `vacuum' solutions, here are not in general playing any dynamical r\^ole. (Indeed we are in Euclidean space.) Their r\^ole arises here only because they generate reparametrisations.

\subsubsection{Other examples}

Vanishing on the equations of motion implies redundancy but it is not an exhaustive diagnostic test. In flat space, one can also consider symmetries based on scaling the coordinates, inducing a change $F \propto x^\mu \partial_\mu\ph(x)$. On substitution in \eqref{redundant}, the explicit dependence on $x$ disappears by integration by parts, but the result is an operator that \emph{does not} vanish on equations of motion. An example of a redundant eigenoperator of this type arises at the next order in the derivative expansion, $O(\partial^2)$, if a power law cutoff is used, as a result of an extra scaling symmetry \cite{Morris:1994ie,Morris:1994jc}. More general diagnostics can be developed to uncover such redundant operators\cite{Morris:1994jc}. The extra scaling symmetry in refs. \cite{Morris:1994ie,Morris:1994jc}  is an exact $t$-independent symmetry of the scaled equations, and thus the redundant operator is exactly marginal in this formulation, \ie has RG eigenvalue $\lambda=0$.

Depending on the form chosen for the exact renormalisation group, wavefunction renormalisation can be associated with an exactly marginal redundant operator whose underlying symmetry is preserved by the approximation (\eg derivative expansion). This ensures quantisation of the wavefunction renormalisation within these approximations. This was in fact the strategy followed in refs.  \cite{Morris:1994ie,Morris:1994jc,Morris:1996xq}, using the extra scaling symmetry above. The r\^ole of the exactly marginal redundant operator related to wavefunction renormalisation, its preservation in different types of exact renormalisation group and relation to the operator considered in subsec. \ref{sec-LPA} has been considered in detail and in more generality in refs. \cite{Osborn:2009vs,Rosten:2010vm,Osborn:2011kw,Bervillier:2013kda}.

\section{Domain of applicability of the redundancy test}
\label{domain}

We stated that in the LPA, a redundant eigenoperator $v(\ph)$ must
satisfy the  equation \eqref{LPAredundant} for some non-singular
function $\zeta(\ph)$. But the question now arises: over what
domain should $\zeta(\ph)$ be non-singular? It is natural to
require that $\zeta(\ph)$ should be non-singular in fact infinitely
differentiable over the same domain ${\cal D}$ as required for the fixed point
solution $V_*(\ph)$ and its eigenperturbations $v(\ph)$, \ie for
all finite real $\ph$ in this case.

On the other hand, we know that generically $V_*(\ph)$ has a singularity for some finite value of $r$ along any complex ray $\ph = r {\rm e}^{i\theta}$, where by `complex ray' we mean a line such that $0<r<\infty$ and $\theta\ne 0,\pi$. This follows from parameter counting. There are already only a discrete set of acceptable values $V_*(0)$, call them $V^0_*$. These correspond to the discrete set of acceptable FP solutions $V_*(\ph)$ to \eqref{LPAFP} (\ie such that they are real and non-singular for all real $\ph$ and satisfy both the boundary conditions $V'_*(0)=0$ and $V_*(\ph)\sim \ph^{d/(d/2-1)}$ \cite{Morris:1994ki,Morris:1998da}). Requiring that a solution to \eqref{LPAFP} exists for a complex ray parameterised by $\theta$, which satisfies the now three conditions $V_*(0)=V^0_*$, $V'_*(0)=0$ and $V_*(\ph)\sim \ph^{d/(d/2-1)}$, over-constrains the equations leading to no solution. (This was checked numerically during the research of ref. \cite{Morris:1994ki}, although not reported.) The same style of argument can be used to show that eigenoperator solutions $v(\ph)$ generically encounter singularities along any complex ray.

One might be tempted nevertheless to try to insist that $\zeta$
should be non-singular, and indeed bounded, after analytic continuation  to all complex
$\ph$, but this is too restrictive since by Louiville's theorem the
only such function is a constant. Therefore, we have to accept that
in general $\zeta$ is unbounded somewhere in the complex $\ph$ plane
(if only at $\ph=\infty$). In view of the complicated singular
behaviour of $V_*$ and $v$ outside the domain ${\cal D}$ in which they are required to be non-singular, the natural choice is therefore to insist that $\zeta$ is also \emph{only} required to be non-singular in ${\cal D}$.

Now consider what happens when we consider truncations to
polynomials in $\ph$. In this case by design $V_*(\ph)$ and
$v(\ph)$ are entire functions of complex $\ph$. However, by comparison to the LPA
results above, this is clearly an artefact of the approximation. It
would therefore be mistaken to insist that the redundancy test
\eqref{LPAredundant} now hold for $\zeta(\ph)$ also defined as an entire
function. The best we can do in this case is
therefore to require \eqref{LPAredundant} hold again for $\zeta$
non-singular only on ${\cal D}$.

Furthermore, for non-trivial $F=\zeta(\ph)+\cdots$, the symmetry
$\delta\ph = F$ generated by the redundant operator, will  in
general be broken by polynomial truncations (since this will not
respect the constraint to a polynomial of some maximal degree nor respect the constraint placed on the particular large order coefficients of the Taylor
expansion that results in the polynomial truncation). This will result
in the redundant operator disappearing from the eigen-operator
spectrum, but again this is clearly an artefact of the
approximation.

Finally we note that the opposite possibility can arise in general also: the appearance of a redundant operator can be an artefact of the approximation, unless we can establish that the associated symmetry exists in the exact equations (as was actually true for the two examples given in subsec. \ref{examples}).

We therefore arrive at the conclusion that in applying redundancy
tests such as \eqref{LPAredundant} the field transformation
$\zeta(\ph)$ should be required to be non-singular over the same
domain ${\cal D}$ as the fixed point solution $V_*(\ph)$ and its
eigenperturbations $v(\ph)$, with no requirement outside this
domain. Approximations can destroy the reparametrisation symmetry
that is responsible for the redundant operator, but equally they
can allow spurious redundant operators to appear unless the
symmetry responsible for the redundancy can be shown to hold in the
exact equations. If the symmetry cannot thus be established, then
the existence of the redundant operator is clearly a signal that we
need to consider less severe truncations or alternative formulations of the flow equation.

\section{Break-down of LPA-type approximations}
\label{LPA-breakdown}

If we are retaining sufficient space-time derivatives in our approximation, then it is always possible to find some solution to the equations of motion $\delta\Gamma_*/\delta\ph(x)=0$, even if this holds only in some neighbourhood of the space-time point $x$.
The redundancy test \eqref{redundant} therefore in this case always provides, actually quite severe, constraints on the form of the eigenoperator $\cO$. We then have the situation already described in sec. \ref{examples} where redundant operators are very rare and only appear as the result of some symmetry of the chosen form of flow equation.

However at the LPA level the situation can be dramatically different. The redundancy test becomes \eqref{LPAredundant}. If there are no `vacuum' solutions $V'_*(\ph)=0$ in the domain $\D$, then \eqref{LPAredundant} can be trivially inverted to find a non-singular $\zeta(\ph) = v(\ph)/V'_*(\ph)$ for {any} eigenoperator $v$. We see that the lack of such a `vacuum' solution in the LPA results in a dramatic degeneration: \emph{all} the eigenoperators become redundant.  Having divided out by reparametrisations, the remaining physical space of perturbations is empty!

Therefore within the LPA-style approximations we find that either redundant eigenoperators are very rare, only appearing as the result of some symmetry of the equations, or we are unlucky with our equations and every eigenoperator becomes redundant. To our knowledge, the first known example of this latter phenomenon occurs not in scalar field theory but in quantum gravity, as we now show.

\section{Redundant operators in the $f(R)$ approximation to quantum
gravity}
\label{f(R)}

\subsection{Preliminaries}

Within the exact renormalisation group framework, several versions
of the $f(R)$ approximation to quantum gravity have been derived
\cite{Machado, Codello,
Benedetti,Demmel:2012ub,Demmel:2013myx,Benedetti:2013jk}. This is,
as Benedetti and Caravelli \cite{Benedetti} have emphasised, as
close as one can get to the LPA in this context.  Very briefly,
the derivation proceeds as follows. The covariant background field
framework is utilised, writing the full metric as  the sum of a
background and quantum (or fluctuation) field: $g_{\mu\nu}={ \bar
g}_{\mu\nu}+h_{\mu\nu}$, the second order functional
differentiation in \eqref{erg} now being performed with respect to
the fluctuation field $h_{\mu\nu}$ \cite{Reuter1}. A gauge choice
is implemented, bringing with it the corresponding ghosts. A
transverse decomposition of the fields is performed, bringing with
it further auxiliary fields.\footnote{In \cite{Demmel:2012ub,Demmel:2013myx}  gravity in $d=3$ dimensions is instead treated in a conformal truncation.} All these fields are given cutoff
profiles that play the r\^ole of $\R$ and incorporate their
contributions into \eqref{erg}. Three types of approximation are
then made. Firstly, $k$ dependence in the ghost \cite{Reuter1} and
auxiliary terms on the left hand side of \eqref{erg} is neglected.
This means that the ghost and auxiliary fields then drop out of the
equations: they only contribute indirectly through the structure of
the right hand side of \eqref{erg}. Secondly, mixed terms depending
on both $h_{\mu\nu}$ and $g_{\mu\nu}$ in the left hand side are
neglected. This means that the $h_{\mu\nu}$ functional derivatives
in \eqref{erg} can be regarded as evaluated at $h_{\mu\nu}=0$, and
afterwards the background field ${\bar g}_{\mu\nu}$ and full metric
$g_{\mu\nu}$ may be identified. We will make this identification
from now on. Finally, the dependence on the metric is truncated to
\be
\label{fRapprox}
\Gamma=\int\!d^dx\,\sqrt{g}\,f(R,t)\,,
\ee
which means that on evaluating the right hand side of \eqref{erg}, only terms that take this general form are kept.

It is this last step which will be of most interest to us. The projection to \eqref{fRapprox} is achieved by working on a maximally symmetric four-dimensional Euclidean space of positive curvature $R>0$, in other words a four-sphere. In this case all the dependence on $g_{\mu\nu}$ necessarily collapses to the form \eqref{fRapprox} since
\be
\label{maximal-symmetry}
R_{\mu\nu} = {R\over d}\,g_{\mu\nu}\,,\qquad R_{\rho\sigma\mu\nu} = {R\over d(d-1)}(g_{\rho\mu}g_{\sigma\nu}-g_{\rho\nu}g_{\sigma\mu})\,,\qquad{\rm and}\qquad \nabla_\mu R = 0\,.
\ee

The flow equations for $f$ derived in refs. \cite{Machado, Codello,
Benedetti} are rather involved but we will not need their explicit
form in what follows. In ref. \cite{Dietz}, we analysed the
properties of the resulting space of fixed point solutions $f_*(R)$, including their eigenoperator spectrum. We showed that no
sensible fixed point solutions $f_*(R)$ to the equations of refs.
\cite{Machado, Codello} exist over the required domain $0\le
R<\infty$. The problem arises from singularities induced by the
choice of cutoff functions. On the other hand in ref.
\cite{Benedetti} careful choices of cutoff functions,  in
particular,  remove many of the singular points. We showed that the
resulting flow equation does support global smooth fixed point
solutions. However now these appear as lines of fixed points, with
each fixed point supporting a continuous spectrum of
eigenoperators.

\subsection{Flow equations of Benedetti and Caravelli}

We will now show that this unexpected behaviour is associated with a break-down of the $f(R)$ approximation in the equations of ref. \cite{Benedetti} analogous to that described in sec. \ref{LPA-breakdown}.
In the $f(R)$ approximation, an eigenoperator takes the form
\be
\label{fR-operator}
\int\!\!d^dx \sqrt{g}\, v(R)\,.
\ee
Similarly to \eqref{reparam}, we take the eigenoperator to be
redundant if it is generated by the change of metric field
\be
\label{gparam}
g_{\mu\nu}(x)\mapsto g_{\mu\nu}(x)+\eps\,  F_{\mu\nu}[g](x)\,,
\ee
to first order in $\eps$,  since   clearly then the flow along this direction in theory space leaves the physics unchanged. The integrated operator then takes
the form
\be
\label{g-redundant}
\int\!\!d^dx \sqrt{g}\, F_{\mu\nu} \frac{\delta\Gamma}{\delta g_{\mu\nu}}\,.
\ee
A redundant
operator in the $f(R)$ approximation therefore takes the form
\be
\label{fR-redundant-general}
\int\!\!d^dx \sqrt{g}\, F_{\mu\nu} \left\{ \half g^{\mu\nu} f_*-
R^{\mu\nu} f'_* + \nabla^\mu\nabla^\nu f'_* - g^{\mu\nu}\, \Box
f'_* \right\}\,.
\ee
However on a constant curvature background the last two terms
vanish, whilst maximal symmetry through \eqref{maximal-symmetry} implies
that $F$ is restricted to the form
\be
\label{delta-g}
F_{\mu\nu} = \zeta(R)\, g_{\mu\nu}\,.
\ee
Therefore in this approximation scheme, an eigenoperator is
redundant if and only if
\be
\label{fR-redundant}
v(R) = \zeta(R)\, E_d(R)
\ee
for some non-singular function $\zeta(R)$, where
\be
\label{eofm}
E_d(R)={d\over2} f_*(R)-Rf'_*(R)\,.
\ee
We see therefore that an eigenoperator $v$ is redundant if and only if it vanishes when $E_d$ vanishes. Like \eqref{LPAredundant}, this corresponds to vanishing on the equations of motion of the `vacuum', where here the r\^ole of the vacuum is played by spaces of constant curvature.

The choice of domain over which the redundancy test \eqref{fR-redundant} applies is crucial. We have already argued in sec. \ref{domain} that we must use only the same domain over which $f_*$ and $v$ are already required to be non-singular. We have checked that for all the fixed point solutions we found in ref. \cite{Dietz} that $E_4$ actually vanishes nowhere in the range $0\le R<\infty$. It follows therefore that for all these fixed points, \eqref{fR-redundant} is  trivially satisfied, and therefore every eigenoperator is redundant. This includes the exactly marginal operator $\delta f_*(R)$ that shifts the fixed point infinitesimally along the line of fixed points. Therefore all the fixed points in any given line of fixed points are equivalent to each other under reparametrisation.

The unexpected results of ref. \cite{Dietz} can therefore be understood as largely due to a dramatic degeneracy of the particular $f(R)$ flow equations \cite{Benedetti} in this domain: factoring out the infinitely many redundant operators which merely reparametrise the fixed point solutions, we are left only with a discrete set of fixed points (a single representative on each line) each of which supports only a zero dimensional space of eigenperturbations. Since there are no perturbations left, there is no real physical sense in which members of this discrete set can be considered different. Actually very likely even members of this discrete set are all equivalent under appropriate finite reparametrisations, \ie in the sense that around any of  these fixed points any other member of the discrete set can be reached by a finite flow along some trajectory specified by starting with some linear combination of the continuously infinite set of redundant operators.

In ref. \cite{Dietz}, we showed that sensible fixed point behaviour could
in principle be recovered by matching smoothly into spaces with
constant negative curvature. One way to do this would be
to analytically continue the equations of ref. \cite{Benedetti} to
negative $R$, but we were not successful in finding global
solutions in this case.
Note that this suggestion can still be consistent with our discovery above that this $f(R)$ approximation has degenerated when restricted to four-spheres.
Indeed if there had been a solution $f_*(R)$ valid for all real $R$, the constant curvature equations of motion $E_4(R)=0$ would then have at least one solution as we argue below. Since we also expect a discrete set of fixed point solutions in this case, each supporting a quantised eigenoperator spectrum \cite{Dietz}, the resulting constraint \eqref{fR-redundant} now over-constrains the equations implying that now none of the eigenoperators are redundant. In principle, as we have seen in sec. \ref{examples}, redundant operators could exist for some symmetry reasons but we can be confident that given the complexity of the equations and the approximations that were used in constructing them that no such hidden symmetries exist.

We see therefore that the physical space of solutions is now qualitatively consistent whether we work only on four-spheres or extend to all real $R$: in both cases we find --up to reparametrisations-- a discrete set of fixed point solutions each supporting a discrete eigenoperator spectrum.

Let us briefly justify the statement above that if a globally well defined  solution $f_*(R)$ to the fixed point equations of ref. \cite{Benedetti} exists over all real $R$, then $E_4(R)=0$ for some real $R$. In ref. \cite{Dietz} we showed that asymptotically
\be
\label{asymp}
f_*(R) = A \,R^2 + R\left\{\frac{3}{2}A+B\cos\ln R^2 + C\sin\ln R^2\right\} + O(1)\,,
\ee
where $A,B,C$ are three real parameters subject only to the constraint that they lie within a cone given by the inequality $\frac{121}{20}A^2 > B^2 + C^2$. Thus asymptotically,
\be
\label{asympE}
E_4(R) = R\left\{\frac{3}{2}A+(B-2C)\cos\ln R^2 + (C+2B)\sin\ln R^2\right\}+O(1)\,.
\ee
This vanishes at an infinite number of points unless $\frac{9}{20}A^2 > B^2 + C^2$. On the other hand if this inequality is satisfied then $E_4(R)$ asymptotically has the same sign as  $AR$; if we can trust that the value of $A$ in \eqref{asymp} is the same for both $R$ positive and negative, then also in this case $E_4(R)$ changes sign and must therefore vanish for some real $R$ by continuity. Analytically continuing \eqref{asymp} into the complex plane, by writing $R=|R|\,{\rm e}^{i\vartheta}$ and considering \eqref{asymp} for increasing $\vartheta$, we see that the domain of $f_*(R)$ is multi-sheeted with $B$ and $C$  taking complex values dependent upon the sheet. However the parameter $A$ is unaffected and therefore does take the same value for both positive and negative large $R$. This completes the demonstration that if a global solution $f_*(R)$ exists defined over all real $R$ then $E_4(R)=0$ also has a solution in this domain.

\subsection{Alternative flow equations of Benedetti}
\label{alternative}

In ref. \cite{Benedetti:2013jk} an alternative $f(R)$ approximation was proposed where the cutoff functions are chosen to be independent of the effective action \eqref{fRapprox}. The fixed point solutions are also now assumed to match smoothly from the sphere ($R>0$) into the hyperboloid ($R<0$). The eigenoperator equations can then be argued to be of Sturm-Liouville type and therefore the RG eigenvalues form a discrete spectrum with finitely many relevant directions \cite{Benedetti:2013jk}. Since the fixed point equation is a second order non-linear ordinary differential equation with two boundary conditions imposed (namely $f_*(R)\sim AR^2$ as $R\to\pm\infty$) we see that it also follows that there is at most a discrete set of fixed point solutions. Although no explicit solutions have so far been attempted, this picture is qualitatively the same as the one we sketched above (with the added information that the number of relevant operators is bounded about any fixed point).
Since there is no need here for the eigenoperators to be redundant (in the sense that we already have a discrete set of fixed points supporting a quantised spectrum)  we predict that if fixed point solutions to the equations of \cite{Benedetti:2013jk} exist, they do allow `vacuum' solutions $E_4(R)=0$ for some real $R$.

\subsection{Polynomial truncations to the equations of Codello, Percacci and Rahmede}
\label{codellos}

Now let us comment on the results derived from polynomial truncations \cite{Reuter1,Niedermaier:2006wt,Percacci:2007sz,Litim:2008tt,Reuter:2012id,Falls:2013bv}. These have recently been taken to very high order \cite{Falls:2013bv} based on expanding the $f(R)$ flow equations in ref. \cite{Codello}. As we remarked already in sec. \ref{domain}, the general reparametrisation symmetries (\ref{gparam},\ref{delta-g}) if they were present in the first place, do not survive such truncations. Given that the results from the truncations are in fact very well behaved \cite{Falls:2013bv},  one might expect that the resulting fixed point solutions $f_*(R)$ do support a vacuum solution $E_4(R)=0$ somewhere within their domain of validity. Truncations automatically explore the analytically extended space, in particular negative curvatures. From ref. \cite{Daniel}, truncations converge over a range $|R|\lesssim0.82$ and match an exact partial solution\footnote{Of course following ref. \cite{Dietz}, the solution cannot be global.} that makes it to the first positive fixed singularity $R_c=2.0065$ \cite{Dietz}. Unfortunately, numerical analysis of this exact but partial solution shows that there are no solutions to $E_4(R)=0$, so here too all eigenoperators are redundant \cite{Kevin}.

\subsection{The Einstein-Hilbert action in general background}
\label{sec-EHGB}

For completeness we finish with some more remarks about perturbations around the Einstein-Hilbert action $f_{EH}(R)$, \cf \eqref{EH}.  For the moment consider general perturbations (rather than restricted to being eigenoperators) and indeed lift the requirement that $f$ corresponds to a fixed point. For this action, whatever the metric, the last two terms vanish in \eqref{fR-redundant-general}. General reparametrisations of the form \eqref{delta-g} therefore result in a perturbation that is of $f(R)$ type: (\ref{fR-operator},\ref{fR-redundant}). At first sight therefore the Einstein-Hilbert action and much more general $f(R)$ actions are equivalent even considered as functionals of an arbitrary metric.\footnote{We thank Kostas Skenderis for this observation, which provided the initial inspiration for the whole paper.} Note however that the perturbation must still factorise on the equations of motion $E_d(R)=0$ (with of course $f_*$ replaced by $f_{EH}$). More importantly the resulting Lagrangian, $f(R)=f_{EH}(R) +\eps\zeta(R)E_d(R)$, is only infinitesimally different from Einstein-Hilbert; as soon as $f(R)$ contains monomials $R^n$ of higher power than $n=1$ with finite coefficients, we cannot neglect the last two terms in \eqref{fR-redundant-general} for general metric, and it is no longer true that reparametrisations \eqref{delta-g} map us to a new Lagrangian of $f(R)$ form. These observations are the equivalent of remarks we made at the beginning of sec. \ref{LPA-breakdown}. Indeed the last two terms of \eqref{fR-redundant-general} are the derivative terms responsible for propagating the physical scalar mode that appears in $f(R)$ gravity.

 On the other hand, working around the Gaussian fixed point $G=\Lambda=0$, it is important to recognise that infinitesimal reparametrisations of the metric can generate arbitrary (positive integer) powers of $R$, more generally any combination that vanishes on shell: $R_{\mu\nu}=0$.  This was the starting point for the perturbative analysis of the obstructions to renormalisability \cite{Goroff-Sagnotti} (where the reparametrisations are all discarded by insisting that $R_{\mu\nu}=0$). Although it underlines why it is important to go beyond $f(R)$ for a better understanding of the ultraviolet behaviour of quantum gravity \cite{beyond-f}, the $f(R)$ approximation still tests an infinite number of physical couplings in this case as we now emphasise.

\subsection{The Einstein-Hilbert action and the Gaussian fixed point on a four-sphere}
\label{EH2}

Returning to the $f(R)$ approximation defined by projecting the background on a four-sphere then,
one might be tempted by the last paragraph to conclude that the $f(R)$ approximation degenerates about the Gaussian fixed point in the sense explained at the beginning of this section, namely that all eigenoperators are then redundant. This would mean that there are no perturbations left to explore renormalisability. However, the reparametrisations all vanish at $R=0$, \ie have $R$ as a factor. On the other hand the eigenoperators generically all have a constant (\ie cosmological constant) piece. This is the cosmological constant problem in a different guise,  ignored in \cite{Goroff-Sagnotti} only because the quartic divergence is invisible in dimensional regularisation.
 We therefore have the standard scenario where redundant operators (which must be eigenoperators \cf sec. \ref{subsec-redundant}) will appear as a result of symmetries within a particular choice of renormalisation group.

\section{Conclusions}
\label{conclusions}

By applying the theory of redundant operators in LPA-style
truncations that we built up in previous sections, a consistent
picture has emerged from sec. \ref{f(R)} for the properties of
non-trivial fixed point solutions $f_*(R)$ found so far in $d=4$
dimensions. Recall that these follow from flow equations which are
derived by projecting on four-spheres. The $f(R)$ flow equations of
ref. \cite{Benedetti} allow fixed point solutions that are globally
well defined on four-spheres (\ie $0\le R<\infty$) \cite{Dietz},
however they appear to have a radically different physical space
from that expected by the now-standard lore
\cite{Reuter1,Niedermaier:2006wt,Percacci:2007sz,Litim:2008tt,Reuter:2012id,Falls:2013bv,Daniel},
forming lines of fixed points in this domain with each fixed point
supporting a continuously infinite spectrum of eigenoperators
\cite{Dietz}. But the solutions we found do not support `vacua'
$R=R_*>0$ there. As a consequence the $f(R)$ approximation
degenerates, with all eigenoperators becoming redundant. There is
no physical space of perturbations; all fixed points on a line are
related by a continuous reparametrisation and actually, as we
argued, physically the whole theory space collapses to a single
point.

In principle we could make progress by extending the domain of validity of the flow equations \cite{Benedetti} to constant negative curvature spaces so that now $R$ spans the whole real line. We showed that the asymptotic behaviour established in ref. \cite{Dietz} implies that if a global fixed point solution $f_*(R)$ still exists, then there are now `vacua' $R=R_*$ within the extended domain. At the same time, we established in ref. \cite{Dietz} that such fixed points would be discrete in number and support a quantised spectrum of eigenoperators. As we saw, it then follows that around these fixed points none of the eigenoperators are redundant. Unfortunately we did not find any global fixed point solution in this enlarged domain \cite{Dietz}.

The flow equations of refs. \cite{Machado,Codello} are prevented from having global fixed point solutions by the appearance of fixed singularities that can be traced to the properties of the cutoff used \cite{Benedetti,Dietz}. As we saw in sec. \ref{codellos}, the partial fixed point solution picked out by polynomial truncations of these equations  \cite{Falls:2013bv,Daniel}, also suffers from a break-down of the $f(R)$ approximation. Although the flow equations of ref. \cite{Benedetti} appear to be an improvement in that sufficiently many fixed singularities are avoided, we have seen that unfortunately the resulting global fixed point solutions are still plagued with this newly discovered unphysical behaviour. To make further progress it is clearly desirable to understand what is the underlying cause. We need to search for flow equations that avoid both of these pitfalls. Perhaps these are already furnished in ref. \cite{Benedetti:2013jk}.

On the other hand, as we have already remarked in subsec. \ref{sec-EHGB},  it is clearly desirable to go beyond the $f(R)$ approximation. As we saw in secs. \ref{LPA-breakdown} and \ref{f(R)}, the break-down where all eigenoperators become redundant seems only possible with these LPA-style truncations.  In principle we can go beyond the $f(R)$ approximation by retaining other higher derivative terms  \cite{beyond-f} and/or more of the action for the quantum field $h_{\mu\nu}$ and ghosts \cite{Eichhorn:2013ug}.
Keeping more of these would allow one to explore the relation between reparametrisations of the quantum field $h_{\mu\nu}$ which, recalling the analysis leading to \eqref{eq-F}, is where the reparametrisations can be regarded as acting, and the extent of its
equivalence to reparametrising the background field $g_{\mu\nu}$, together with the constraints that arise from modified Ward identities and the ghost action.

\section*{Acknowledgments}
It is a pleasure to thank Kostas Skenderis and Dario Benedetti for initial discussions, and Kevin Falls, Roberto Percacci, Claude Bervillier and one of the referees for useful comments on the first version of the paper. TRM acknowledges STFC (CG ST/J000396/1) for financial support.

\end{document}